\documentclass{phc-proc4-auth}

\usepackage[centertags]{amsmath}
\usepackage{amsfonts}
\usepackage{amssymb}
\usepackage{newlfont}
\usepackage{exscale}
\usepackage[dvips]{graphicx}
\usepackage{subfigure}
\usepackage{epsfig}
\usepackage{floatflt}
\usepackage{amsmath}

\begin{document}
\begin{frontmatter}
\title{Resonant Raman scattering in mercurate single crystals}
\author{Y. Gallais$^1$, A. Sacuto$^{1,2}$ and D. Colson$^3$}
\address{$^1$ Laboratoire de Physique du Solide (UPR 5 CNRS) ESPCI, \\
10 rue Vauquelin 75231 Paris. \\
$^{2}$ Mat\'eriaux et Ph\'enom\`enes Quantiques (FER 2437 CNRS),
Universit\'e Paris 7 \\
2 place Jussieu 75251 Paris.\\
$^{3}$ Service de Physique de l'Etat Condens\'{e}e,
CEA-Saclay,91191 \\
Gif-sur-Yvette, France}
\maketitle

\begin{abstract}
We report resonant electronic Raman scattering in optimally doped
single layer HgBa$_2$CuO$_{4+\delta}$ (Hg-1201) and trilayer
HgBa$_2$Ca$_2$Cu$_3$O$_{8+\delta}$ (Hg-1223) single crystals.
Analysis of the $B_{1g}$ and B$_{2g}$ channels in the
superconducting state of Hg-1201 advocates for a gap having d-wave
symmetry. In addition a resonant study $B_{1g}$ pair-breaking peak
and the $A_{1g}$ peak suggests that the $A_{1g}$ peak is not
directly related to the d-wave superconducting gap amplitude.
Comparison with trilayer Hg-1223 demonstrates the universal
behavior of this two energy scales in optimally doped cuprates.
\end{abstract}
\end{frontmatter}

\section{Introduction}

In recent years electronic Raman scattering (ERS) has been used
extensively to probe both the normal and the superconducting
states of the cuprates. While the broad Raman continuum in the
normal state is only poorly understood at present, Raman spectra
in the superconducting state have been described quite
successfully in terms of BCS theory with a d-wave gap in optimally
doped cuprates \cite{dev94,sacuto2000}. By using different
incident and scattered light polarisation ERS probes different
directions in reciprocal space allowing an analysis of the gap
symmetry. The $B_{1g}$ symmetry is sensitive to the $k_x=0$ and
$k_y=0$ directions and thus gives a pair-breaking peak at
$2\Delta_0$ while the $B_{2g}$ symmetry probes the $k_x=k_y$
directions and is sensitive to the nodes of the d-wave gap.
Although spectra taken in these two symmetries cuprates are
consistent with a d-wave gap in most optimally doped cuprates,
spectra taken in the fully symmetrical A$_{1g}$ channel reveal a
peak below $2\Delta_0$ whose intensity and position are difficult
to reconcile with a simple d-wave gap picture \cite{strohm}. In
this symmetry, vertex corrections due to Coulomb interaction and
charge conservation (or backflow) are expected to strongly
suppress the scattering intensity in the superconducting state
\cite{klein-dierker}. Therefore the intensity in the $A_{1g}$
channel is expected to be much smaller than in the $B_{1g}$
symmetry. This contradicts experimental results in most cuprates
which show an $A_{1g}$ peak intensity stronger than the $B_{1g}$
one in most cases \cite{dev94,sacuto2000,gallais}. Many scenarios
have been proposed to explain this discrepancy: coupling between
adjacent CuO$_2$ planes \cite{strohm}, resonant effect of the
Raman vertex \cite{sherman} and, for three and four layer
cuprates, contribution from the c-axis bilayer plasmon observed in
the far-infrared spectra of the c-axis conductivity \cite{munzar}.
Recently we have shown using Ni substitution in optimally doped
Y-123 that the energy of the $A_{1g}$ peak follows intriguingly
the energy of the neutron resonance detected in the spin
excitation spectrum by Inelastic Neutron Scattering (INS)
\cite{gallais}. However the exact microscopic mechanism of the
$A_{1g}$ peak remains unclear and more experimental work is
clearly needed. In this paper we report Raman scattering in
HgBa$_2$CuO$_{4+\delta}$ (Hg-1201) and
HgBa$_2$Ca$_2$Cu$_3$O$_{8+\delta}$ (Hg-1223) where the $2\Delta_0$
pair breaking peak and the A$_{1g}$ peak have been studied as a
function of the excitation energy. Both energy scales show very
different resonant behaviors in Hg-1201, suggesting different
underlying microscopic mechanisms. Comparison with the trilayer
Hg-1223 demonstrates that the nature of the two energy scales is
independant of the number of CuO$_{2}$ layers and is therefore a
universal property of the superconducting CuO$_{2}$ plane.

\section{Experimental}

Raman scattering experiments were performed on two different
single crystals: the single layer Hg-1201 (T$_c$=95~K) and the
trilayer Hg-1223 (T$_c$=125~K). Both crystals are very close to
optimal doping and have tetragonal symmetry. The single crystal of
Hg-1201 has been successfully grown by the flux method. The
detailed procedure for crystal growth will be described elsewhere
\cite{colson03}. Details on the synthesis of Hg-1223 single
crystals can be found in ref. \cite{colson94}. Plane polarised
spectra were taken using different lines from a Ar$^+$-Kr$^+$
laser and by a JY T64000 spectrometer (except for the Hg-1223 data
where a JY U1000 was used). The spectra were corrected from the
spectrometer response, the Bose factor and the optical constants
thus yielding the imaginary part of the Raman response.
Temperature indicated are corrected from the laser heating for
Hg-1201 spectra (typically 10 to 15K) and are the nominal
temperatures for Hg-1223 spectra.

\section {Results and discussion}

\subsection{Hg-1201}
\begin{figure}

\centering \epsfig{figure=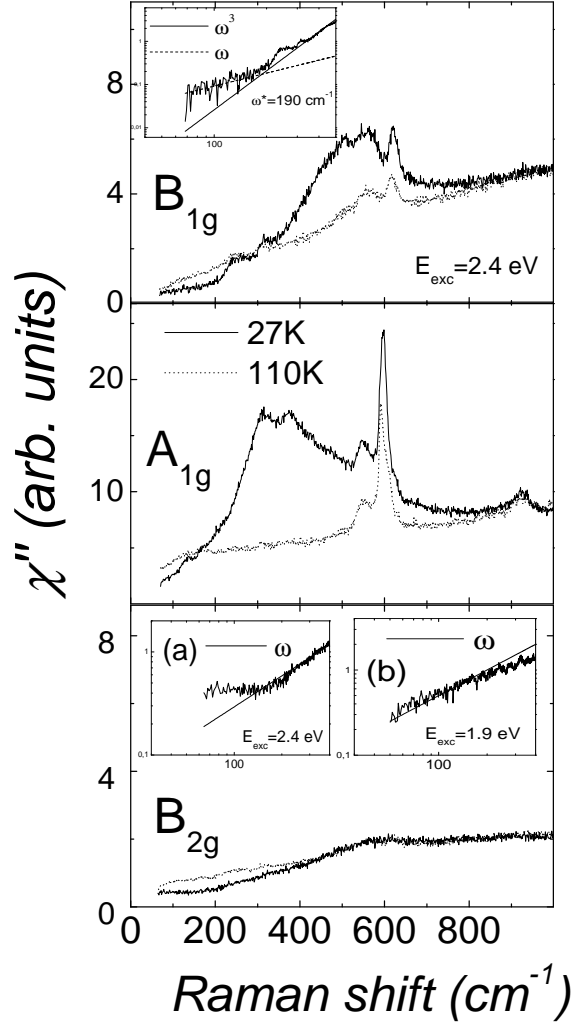, width=0.98\linewidth}
\caption{Raman responses of optimally doped Hg-1201 in $B_{1g}$,
$A_{1g}$ and $B_{2g}$ symmetries in the normal and superconducting
states. The $A_{1g}$ symmetry was obtained by substracting the
$B_{2g}$ component to the spectra taken in (x'x') scattering
geometry ($A_{1g}$ + $B_{2g}$). Insets show log-log plots of the
response in $B_{1g}$ and $B_{2g}$ symmetries with linear and cubic
fits. All spectra are taken using the 2.4~eV excitation line
except spectrum shown in the inset (b) for the $B_{2g}$ channel
which was performed using the 1.9~eV excitation line.} \label{514}

\end{figure}

In Figure \ref{514} we show spectra in the $B_{1g}$, $A_{1g}$ and
$B_{2g}$ channels in the normal and superconducting states using
the 2.4~eV (514.5~nm) excitation line. In the $B_{1g}$ channel,
which probes the antinodal region (($\pi$,0) points), the response
in the superconducting state shows the usual $2\Delta_0$ pair
breaking peak located around 520~cm$^{-1}$. This gives a coupling
ratio $\frac{2\Delta_0}{k_BT_c}$ of 7.9.  The response in the
$A_{1g}$ symmetry on the other hand shows a very strong
superconductivity induced peak located below the $2\Delta_0$ at an
energy of 330~cm$^{-1}$ ($\frac{E_{A_{1g}}}{k_BT_c}$=5.1). The
presence of the two different energy scales is consistent with
data on other optimally doped cuprates \cite{gallais}. We note
that the presence of a strong $A_{1g}$ peak in a single layer
cuprate rules out a scenario in which the peak is due to
unscreened charge fluctuation between adjacent CuO$_2$ layers. In
the $B_{2g}$ symmetry, which probes the nodal region, we do not
observe any superconducting induced peak. The presence of a weak
peak in the superconducting state of the $B_{2g}$ symmetry appears
to be material dependant and could be sensitive to disorder
\cite{dev95}.

\begin{figure} \centering
\epsfig{figure=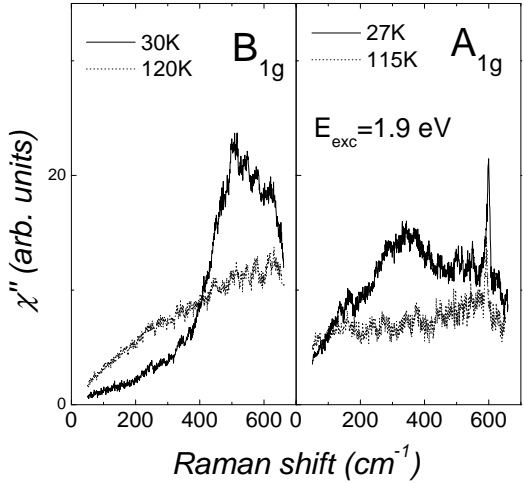, width=0.9\linewidth} \caption{Raman
responses of Hg-1201 in the $B_{1g}$ and $A_{1g}$ symmetries using
the 1.9~eV excitation line.} \label{647}
\end{figure}

\begin{figure}
\centering \epsfig{figure=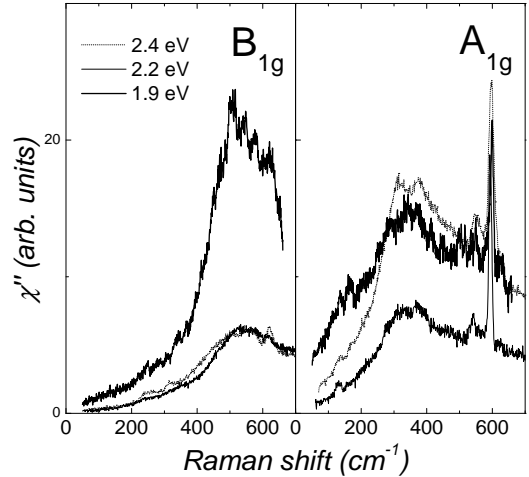, width=0.9\linewidth}
\caption{$B_{1g}$ and $A_{1g}$ symmetries in the superconducting
state (T=27K-30K) taken at three different excitation energies.}
\label{resonance}
\end{figure}

As was pointed out by Devereaux et al. \cite{dev94} careful
inspection of the power law behaviors in the $B_{1g}$ and $B_{2g}$
channels can yield information on the gap anisotropy. For a clean
d-wave superconductor the low energy Raman response in $B_{1g}$
symmetry is expected to follow a cubic law while the $B_{2g}$
symmetry should be linear. In the $B_{1g}$ channel a crossover
between linear and cubic behavior is observed around 190~cm$^{-1}$
(see inset) consistent with theoretical prediction for disordered
d-wave superconductors \cite{dev95}. Similar behavior in this
symmetry using the 2.2~eV and the 1.9~eV excitation is observed
(see Figures \ref{647} and \ref{resonance}). On the other hand
spectrum in the $B_{2g}$ is linear except below 140~cm$^{-1}$
where additional scattering is found producing a nearly flat
continuum at low energy. The origin of this anomalous additional
intensity is unclear but the same spectrum taken using the 1.9~eV
excitation energy display a linear behavior at low energy down to
the lowest energy measured (50~cm$^{-1}$) suggesting the presence
of nodes in the ($\pi$,$\pi$) direction in the gap function (see
inset (b) of Figure \ref{514}). Thus the increase of scattering at
low energy using the 2.4~eV excitation is most likely spurious and
we conclude that the present data indicate a gap of d-wave
symmetry in optimally doped Hg-1201.

In figures \ref{647} and \ref{resonance} we examine the resonant
behavior of the $B_{1g}$ and $A_{1g}$ peaks. In Figure \ref{647}
we have plotted spectra in the normal and superconducting states
performed using the 1.9~eV excitation. We note that if both peak
are located at the same energy as the one taken using the 2.4~eV,
their relative intensities have changed drastically.
While for the 2.4~eV excitation the $A_{1g}$ peak intensity more
than twice the $B_{1g}$ pair breaking peak intensity, the
situation is opposite in the case of the 1.9~eV excitation energy
where the $A_{1g}$ peak intensity is now a factor 1.5 smaller than
the $B_{1g}$ pair breaking peak. In Figure \ref{resonance} the
$B_{1g}$ and $A_{1g}$ channels in the superconducting state are
shown as a function of the excitation energy used. Similarly to
the case of Y-123 and Bi-2212 the $A_{1g}$ peak position in
Hg-1201 is found to be insensitive to the choice of excitation
energy and remains centered around 330~cm$^{-1}$. Its intensity is
nearly constant between 2.4~eV and 1.9~eV but is decreased by
about a factor 2 when using the 2.2 eV excitation line (see Figure
\ref{resonance}). The $B_{1g}$ pair breaking peak on the other
hand shows a remarquable resonant behavior toward 1.9~eV (see
figure \ref{resonance}): its intensity is nearly the same for
2.4~eV and 2.2~eV but is enhanced for 1.9~eV by more than a factor
4. In a similar way the normal state $B_{1g}$ continuum intensity
is also enhanced towards 1.9 eV (see Figures \ref{514} and
\ref{647}). Such a simultaneous resonant behavior in both the
normal and superconducting states in the $B_{1g}$ channel has also
been observed in Tl2201 \cite{kang} and strongly advocates for an
electronic origin of the normal state $B_{1g}$ Raman continuum
below 1000~$cm^{-1}$.

Previous resonant Raman scattering studies have been mainly
focused on the resonant properties of the $B_{1g}$ (or $B_{2g}$)
pair breaking peak. Resonant Raman scattering data in Tl-2201
\cite{kang} show a strong increase of the $B_{1g}$ peak intensity
toward high excitation energies (3~eV) and more recently resonant
behavior of the B$_{2g}$ channel towards low excitation energy
(1.9 eV) was found in electron doped NCCO \cite{blumberg}.
In a free electron-like picture of Raman scattering, resonant
behavior is achieved when the incident (or scattered) photon
energy equals an interband distance in k-space. In principle
resonance effects can strongly affect the actual k-dependance of
the Raman vertices and thus be used to selectively probe different
directions in k-space \cite{sherman}. In our case the resonance
enhancement at 1.9~eV in the $B_{1g}$ channel indicates the
presence of an interband transition near the ($\pi$,0) point of
the Brillouin zone in Hg-1201. If we believe band structure
caculations, this transition most likely takes place between the
non bonding bands and the Fermi level (antibonding band)
\cite{Hg-1201-bande}.

We now discuss on the implications of the different resonant
behaviors of the $B_{1g}$ pair breaking peak and the $A_{1g}$
peak.
As mentionned in the introduction, theoretical Raman responses in
a standard BCS d-wave framework systematically fail to reproduce
the experimentally observed $A_{1g}$ response. Probing essentially
the whole Fermi surface the $A_{1g}$ channel is expected to be
greatly influenced by resonances. As a consequence, the $A_{1g}$
peak position, if related to the amplitude of the superconducting
gap, should be very sensitive to the choice of excitation energy,
especially near resonances. In the present case, the resonant
behavior of the Raman vertex in the ($\pi$,0) direction should
result in a shift of the $A_{1g}$ peak position towards
2$\Delta_0$ \cite{sherman}. In contrast our data show that its
position is constant despite the resonance effect in the ($\pi$,0)
direction observed in the $B_{1g}$ channel with the 1.9 eV
excitation line. Thus the $A_{1g}$ peak is a true energy scale
which is not directly related the amplitude of the d-wave
superconducting gap. In fact near the optimally doped regime of
many cuprates this energy scale tracks closely that of the neutron
resonance observed in the spin excitation spectrum by INS
\cite{gallais}.
\subsection{Hg-1223}
\begin{figure}

\centering \epsfig{figure=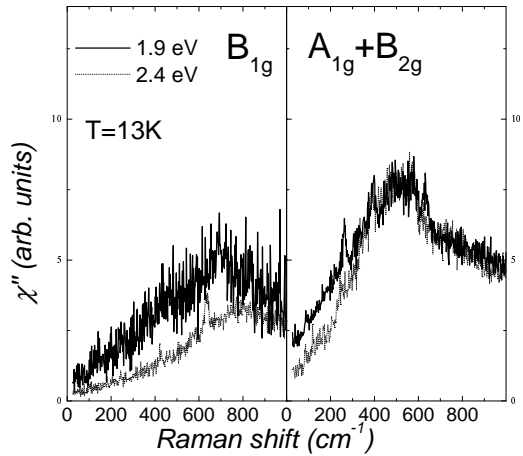, width=0.9\linewidth}
\caption{Raman spectra of Hg-1223 in the superconducting state in
$B_{1g}$ and $A_{1g}$+$B_{2g}$ symmetries at 1.9 and 2.4eV. The
$B_{2g}$ componant is much weaker than the $A_{1g}$ so that the
$A_{1g}$+$B_{2g}$ spectrum is essentially dominated by the
$A_{1g}$ component.} \label{1223}

\end{figure}

 In figure \ref{1223} the $B_{1g}$ and $A_{1g}$
spectra in the superconducting state of Hg-1223 are shown for the
1.9 and 2.4~eV excitation lines. As in the case of Hg-1201 the
$B_{1g}$ response shows a resonance toward 1.9~eV but this time
its intensity increases only by less than a factor 2. The pair
breaking peak position shifts slightly towards lower energy: from
2$\Delta_0$=780~cm$^{-1}$=9.0~k$_BT_c$ for 2.4~eV it moves down to
2$\Delta_0$=720~cm$^{-1}$=8.3~k$_BT_c$ for 1.9~eV. These coupling
ratios are very similar to the one measured in Hg-1201
(2$\Delta_0$=8.1~k$_BT_c$) and suggest that the amplitude of the
gap is controlled by $T_{cmax}$ in the Hg-based cuprates family. A
similar behavior was found in a ARPES study of the Bi family
\cite{sato}. In the case of the $A_{1g}$ channel the peak is
located at 500~cm$^{-1}$=5.7~k$_BT_c$ (5.1~$k_BT_c$ for Hg-1201).
We also note that its intensity is the same for both excitation
energies (again similar to the case of Hg-1201, see Figure
\ref{resonance}). The very similiar behavior between n=1 and n=3
in the $A_{1g}$ channel with respect to both their resonance
behavior and their scaling with $T_c$ rules out the possibility
that the peak is dominated by a c-axis plasmon component in three
and four layers cuprates as has been suggested recently
\cite{munzar}. On the contrary it clearly indicates that the
$A_{1g}$ peak has the same origin for both single and trilayer
cuprates.

\section{Conclusion}

We have reported resonant Raman scattering in single layer
Hg-1201. Analysis of the low energy part of the spectra in the
superconducting indicates a gap of d-wave symmetry. Spectrum in
the $B_{1g}$ channel, contrary to the $A_{1g}$ channel, shows a
remarkable resonant behavior of the pair breaking towards the 1.9
eV excitation energy. The different resonant behavior between both
channels indicates different microscopic origins for the $B_{1g}$
pair breaking peak and the $A_{1g}$ peak. The latter appears to be
related to the neutron resonance energy scale. In addition, data
on trilayer Hg-1223 demonstrate the universal character of the two
energy scales, irrespective to the number of CuO$_{2}$ layers. The
presence in the $A_{1g}$ Raman spectra of an energy scale also
seen in the spin excitation spectrum by INS is presently not
understood and recquires further experimental and theoretical
investigations.

\end{document}